\documentclass[pdflatex,sn-nature,longbibliography,iicol]{sn-jnl}

\usepackage{graphicx}%
\usepackage{multirow}%
\usepackage{multicol}
\usepackage{amsmath,amssymb,amsfonts}%
\usepackage{amsthm}%
\usepackage{mathrsfs}%
\usepackage[scr=rsfs]{mathalpha}%
\usepackage[title]{appendix}%
\usepackage{xcolor}%
\usepackage{textcomp}%
\usepackage{manyfoot}%
\usepackage{booktabs}%
\usepackage{siunitx}
\usepackage{physics} 
\usepackage{chemformula}
\usepackage{algorithm}%
\usepackage{algorithmicx}%
\usepackage{algpseudocode}%
\usepackage{listings}%
\usepackage{tikz}
\usepackage{tikz-3dplot}
\usepackage{pgfplots}
\usepackage{caption}
\usepackage{csvsimple}
\usepackage{geometry}
\usepackage{pgfmath}
\usepackage[inkscapelatex=false]{svg}
\usepackage{subcaption}
\usepackage{lipsum}
\usepackage{natbib}
\usepackage[export]{adjustbox}
\usepackage{titlesec}
\usepackage{etoolbox}
\usepackage{tabularx}
\usepackage[columnwise,switch]{lineno}

\pgfplotsset{compat=1.18}
\theoremstyle{thmstyleone}%
\theoremstyle{thmstyletwo}%

\theoremstyle{thmstylethree}%

\AtBeginDocument{\RenewCommandCopy\qty\SI}

\makeatletter
\newcommand{\mydate}[1]{\gdef\@mydate{#1}}
\let\@mydate\@empty

\patchcmd{\@maketitle}
  {\ifx\authemail\@empty\else Contributing authors:\ \authemail\fi} %
  {\ifx\authemail\@empty\else Contributing authors:\ \authemail\fi %
   \ifx\@mydate\@empty\else\par\vspace{8pt}\centerline{\normalfont\@mydate}\fi}
  {}{}
\makeatother

\begin{document}
\title[Article Title]{Experimental realisation of topological spin textures in a Penning trap}

\renewcommand{\thesection}{\Roman{section}}

\titleformat{\section}
{\normalfont\normalsize\bfseries\centering}
  {\thesection.}{1em}{\MakeUppercase}

\renewcommand{\thesubsection}{\Alph{subsection}.}
\titleformat{\subsection}
  {\normalfont\normalsize\bfseries\centering}
  {\thesubsection}
  {0.5em}
  {}

\setlength{\textwidth}{180mm}
\setlength{\columnsep}{5mm}
\setlength{\oddsidemargin}{-10.4mm}
\setlength{\evensidemargin}{-10.4mm}

\author*[1]{\fnm{Julian Y. Z.} \sur{Jee}}\email{jjee9913@sydney.edu.au}

\author[1]{\fnm{Nihar} \sur{Makadia}}

\author[1]{\fnm{Joseph H.} \sur{Pham}}

\author[1]{\fnm{Gustavo} \sur{Caf\'e de Miranda}}

\author[1]{\fnm{Michael J.} \sur{Biercuk}}

\author[2,3]{\fnm{Athreya} \sur{Shankar}}

\author[1,4]{\fnm{Robert N.} \sur{Wolf}}

\affil[1]{\orgdiv{School of Physics}, \orgname{University of Sydney}, \city{Sydney}, \postcode{2006}, \state{New South Wales}, \country{Australia}}

\affil[2]{\orgdiv{Department of Physics}, \orgname{Indian Institute of Technology Madras}, \orgaddress{\city{Chennai}, \postcode{600036}, \state{Tamil Nadu}, \country{India}}}

\affil[3]{\orgdiv{Center for Quantum Information, Communication and Computing}, \orgname{Indian Institute of Technology Madras}, \orgaddress{\city{Chennai}, \postcode{600036}, \state{Tamil Nadu}, \country{India}}}
\date{}

\affil[4]{\orgdiv{Sydney Nano Institute}, \orgname{University of Sydney}, \orgaddress{\city{Sydney}, \postcode{2006}, \state{New South Wales}, \country{Australia}}}

\abstract{Topological spin textures provide a route to study non-trivial mappings between real-space structure and spin configurations, but their controlled realisation and direct characterisation in large programmable quantum systems remain challenging. Here, we show the deterministic generation and site-resolved reconstruction of topological spin textures in a two-dimensional crystal of more than 150 trapped ions. Using globally applied spin-dependent forces, we engineer spatially varying spin rotations that prepare a skyrmion configuration and reconstruct the full vector spin field from single-ion resolved measurements. Spatially resolved local control further enables the preparation of domain-wall states. Our work establishes trapped-ion crystals as a platform for engineering complex spin textures, providing a foundation for future studies of topology-dependent spin dynamics in programmable quantum systems.}






\maketitle

\begin{figure*}[!htbp]
    \centering
    \includegraphics[width=\textwidth]{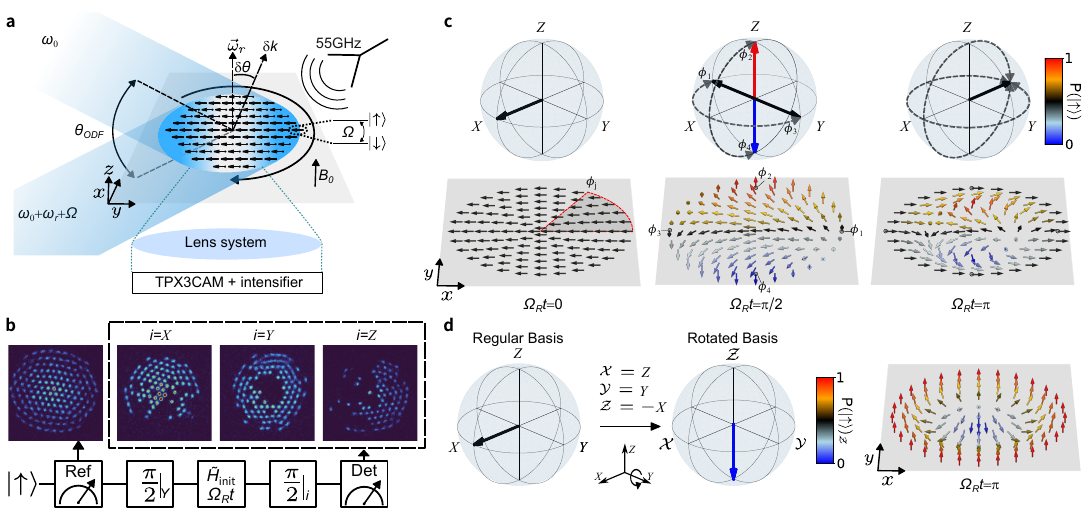}
    \caption{\textbf{Experimental setup and skyrmion initialisation.} %
    (\textbf{a}) Schematic overview of the experiment using a 2D ion crystal, rotating at $\omega_r$ in the lab frame due to a magnetic field $B_0$ along $z$. Global coherent spin rotations, $\ket{\downarrow} \leftrightarrow \ket{\uparrow}$, are driven with microwaves near \qty{55}{\giga\hertz} at Rabi rate $\Omega$. Two off-resonant lasers at optical frequencies of $\omega_0$ and $\omega_0+\omega_r+\Omega$ cross the ion crystal at an angle $\theta_\mathrm{ODF}$ to create a spin-dependent optical dipole force (ODF). Tilting the ODF difference wavevector $\delta \textbf{k}$ by $\delta \theta$ relative to the crystal-plane normal produces a travelling intensity gradient in the $x-y$ plane. Spin-state-dependent fluorescence is captured by a single-photon-timestamping detector (TPX3CAM) capable of resolving individual ions. %
    (\textbf{b}) Pulse sequence for preparing the skyrmion spin texture. Ions are initialised to $\ket{\uparrow}$ and a reference image is acquired to determine their positions. Next, all spins are rotated to the Bloch-sphere equator with a $\pi/2|_Y$ pulse and evolved under $\tilde{H}_\mathrm{init}$ for a duration of $t=\pi/\Omega_R$. Measurements along $i\in\{X,Y,Z\}$ are used to reconstruct the spin texture. The reference image and the corresponding experimental images are shown above the pulse sequence diagram for a crystal containing about 160 ions. %
    (\textbf{c}) Spin state evolution under $\tilde{H}_\mathrm{init}$ is shown with texture arrows representing the nominal spin orientation of each ion. The colour scale represents the probability of measuring a spin in $\ket{\uparrow}$. The target spin texture is realised at $\Omega_Rt=\pi$, where the outermost spins undergo a $\pi$-rotation. Bloch spheres show the trajectories of spins at azimuthal angles $\phi_1$--$\phi_4$, marked on the textures below.%
    (\textbf{d}) The skyrmion spin texture at $\Omega_Rt=\pi$ in the rotated basis. The Bloch spheres show the central spin before and after the frame transformation, with colour indicating the corresponding $Z$- and $\mathcal{Z}$-component, respectively.}
    \label{fig1:experimental_implementation}
\end{figure*}

\section*{Introduction}\label{sec1}

Quantum simulation offers a way to investigate complex interacting systems whose dynamics are often difficult to solve using classical methods. In particular, quantum quenches provide a powerful probe of nonequilibrium dynamics and have been extensively explored in one-dimensional systems \cite{Zhang2017, Jurcevic2017, Tan2021, Dumitrescu2022, De2025}. Extending these studies to two-dimensional (2D) systems is expected to reveal significantly richer behaviour, but remains experimentally challenging. Recent advances using programmable Rydberg atom arrays have begun to access this regime \cite{Ebadi2021, Scholl2021, Manetsch2025}.

Alongside these advancements, trapped-ion quantum simulators offer a promising route to exploring 2D quantum coherent dynamics, combining long coherence times with highly programmable spin-spin interactions and high-fidelity measurements \cite{Blatt2012, Monroe2021}. In particular, Penning traps are promising for large-scale simulations, as they can confine hundreds of ions in 2D laser-cooled Coulomb crystals \cite{Britton2012, Bohnet2016}, while related approaches in Paul traps have recently begun to realise two-dimensional ion arrays \cite{Donofrio2021, Kiesenhofer2023, Guo2024, Sun2024, Qiao2024}.
Recent Penning-trap demonstrations with tens to hundreds of ions have predominantly relied on global measurements and operations, utilising laser-induced forces to create effective interactions between the ions’ internal spin states by coupling them to the crystal's shared collective motion \cite{Garttner2017, Safavi-Naini2018, Bullock2026}. The use of approximately identical coupling of all ions to the collective mode predominantly confines the dynamics to the permutation-symmetric subspace, preventing deterministic preparation of spatially structured spin configurations.

In this work, we go beyond the experimental limitations faced by previous experiments with 2D trapped-ion systems by transforming the rigid-body rotation of the ion crystal in the Penning trap itself into a controllable resource. By precisely tilting the wavefront of a spin-dependent optical-dipole force (ODF) \cite{Pham2024}, we create a spatially dependent coupling that naturally breaks permutation symmetry. This mechanism enables the deterministic preparation of non-uniform and topologically structured spin textures involving hundreds of ions. We further extend the control with a repositionable, tightly focused laser for spatially resolved local control, and utilise site-resolved fluorescence imaging for full spatial readout of the individual spins \cite{McMahon2024, Wolf2024}. Using this system, we experimentally prepare and characterise two distinct spatially structured spin textures: a skyrmion and a domain wall. These textures are prepared as product states that represent distinct initial conditions for quantum quench dynamics in chiral $p$-wave systems and give rise to different subsequent dynamical phases \cite{Foster2013, Lewis-Swan2021, Shankar2022}. More broadly, the methods demonstrated here establish a general framework for preparing, controlling, and characterising complex spin textures in large trapped-ion crystals. 

\begin{figure*}[!htbp]
    \centering
    \includegraphics[width=1\textwidth]{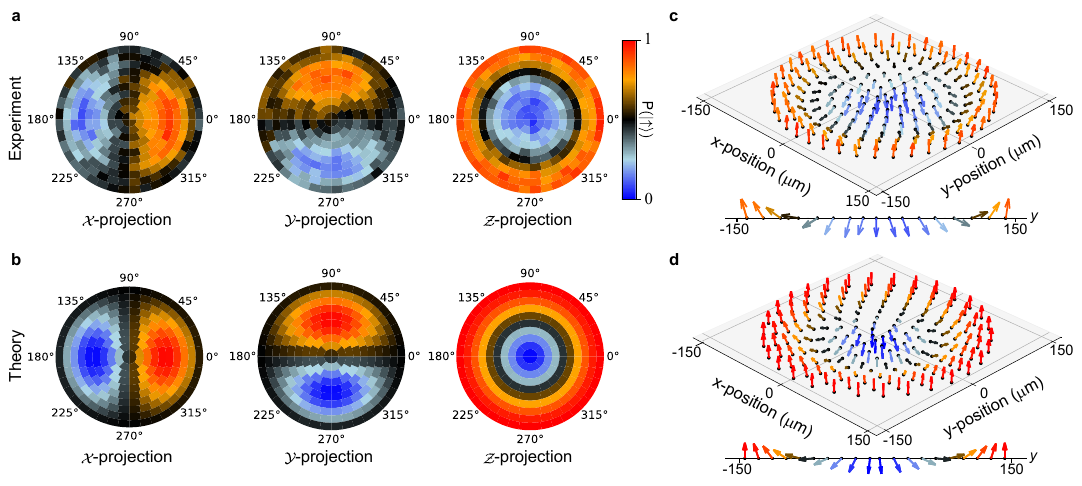}
    \captionof{figure}{\textbf{Experimental reconstruction of the skyrmion spin texture.} %
    (\textbf{a}) Experimental and (\textbf{b}) theoretical probability distributions of the skyrmion spin texture taken along the $\mathcal{X},\mathcal{Y}$, and $\mathcal{Z}$ bases. To resolve the spatial structure, ions are grouped into $220$ polar bins. Experimental values are averaged over 200 repetitions per projection. The $\mathcal{X}$ and $\mathcal{Y}$ projections reveal characteristic dipole patterns, a signature of the azimuthal spin winding, while the $\mathcal{Z}$ projection shows a radially symmetric gradient. %
    The reconstructed (\textbf{c}) experimental and (\textbf{d}) theoretical spin textures show high qualitative agreement. The plotted vertical cross-section around $x=0$ illustrates the inversion of the $\mathcal{Z}$ component as a function of radius.}
    \label{fig2:spin_texture_reconstruction}
\end{figure*}

\section*{Results}

\subsection*{Skyrmion texture initialisation}\label{sec2}
%
The skyrmion is a radially symmetric, two-dimensional spin texture in which the transverse spin components wind azimuthally, while the longitudinal component varies smoothly with radius. To experimentally realise this configuration, we map the spin texture onto a 2D Coulomb crystal of \ch{^9Be+} ions confined by a Penning trap as shown in Fig. \ref{fig1:experimental_implementation}a. 

A static quadrupolar electric potential superimposed with a \qty{2}{\tesla} magnetic field confines the ions, producing an axial centre-of-mass (COM) mode at \(\omega_\mathrm{COM}/2\pi = \qty{698}{\kilo\hertz}\). The magnetic field Zeeman-splits the valence-electron ground-state manifold by \qty{55}{\giga\hertz}. Within this manifold, we define \(\ket{\uparrow}\) as the ``bright'' state, which is coupled to a closed cycling transition near \qty{313}{\nano\metre} for resonant fluorescence detection and Doppler cooling, and \(\ket{\downarrow}\) as the non-scattering ``dark'' state \cite{Ball2019}. Global initialisation into \(\ket{\uparrow}\) is achieved using \qty{313}{\nano\metre} optical pumping.

The combined force of the trapping electric and magnetic fields induces an $\textbf{E} \times \textbf{B}$ drift, causing the ion crystal to rotate within the trap. The rotation frequency, $\omega_r/2\pi$, can be controlled by an azimuthal rotating quadrupole potential, known as a ``rotating wall" \cite{Huang1998}. This rotation complicates site-resolved spin-state detection with conventional frame-based imaging, as it requires stroboscopic ion-crystal illumination, which is inefficient \cite{Mitchell2001}.

We overcome these limitations using a spatially resolved single-photon timestamping detector, which records the arrival time and position of individual photons \cite{Nomerotski2017} and enables continuous imaging of crystals containing hundreds of ions \cite{Mitchell2001, Bohnet2016}. By reconstructing ion positions in the co-rotating frame, we achieve single-shot localisation and spin-state discrimination without stroboscopic averaging \cite{Wolf2024}. This allows direct reconstruction of spatial spin textures from site-resolved measurements, as shown in Figure \ref{fig1:experimental_implementation}b.

To create the skyrmion texture, we utilise a spin-dependent ODF whose spatial variation is modulated by the rigid-body rotation of the crystal. The ODF is produced by two off-resonant beams whose interference generates a beat note frequency $\mu_r/2\pi$, typically tuned near the axial COM mode to mediate spin-motion coupling. Here, we instead introduce a controlled spatial variation in the coupling by slightly tilting the ODF wavefront relative to the crystal plane. This couples the spins to the crystal rotation, breaking permutation symmetry and imprinting a position-dependent phase across the ions. The relative beam geometry is controlled using optomechanics with millidegree precision~\cite{Pham2024}, enabling precise adjustment of the ODF beam angle $\theta_\mathrm{ODF}$. The relative optical phase is actively stabilised to ensure a reproducible interference pattern (see Methods). The ODF beams intersect at an opening angle $\theta_\mathrm{ODF}\approx\qty{18}{\degree}$, with the difference wave vector tilted by $\delta \theta=\qty{0.04}{\degree}$ from the plane normal. Tilting the ODF difference wavevector $\delta \textbf{k}$ towards $x$ generates an axial component $\delta k_z = |\delta \textbf{k}| \cos(\delta\theta)$ and a radial component $\delta k_x = |\delta \textbf{k}| \sin(\delta\theta)$. For a crystal of radius $R=\qty{150}{\micro\meter}$, this corresponds to an effective in-plane Lamb–Dicke parameter of $\eta_x=\delta k_xR=0.66$.

The experimental protocol used to implement the skyrmion spin texture is outlined in Fig.~\ref{fig1:experimental_implementation}b and is realised by a simultaneous application of the ODF and a global microwave drive (see Appendix~\ref{sec_Hamiltonian_derivation}). This is described by the Hamiltonian (setting $\hbar=1$)

\begin{multline}
\label{eqn:H_init}
H_{\mathrm{init}} =
\sum_j \frac{\omega_s}{2}\hat{\sigma}_j^Z 
+ \sum_n \omega_n \hat{a}^\dagger_n \hat{a}_n+
\sum_j \Omega\cos\left(\omega_s t\right)\hat{\sigma}_j^X\\
+\sum_j \delta_{\mathrm{ac}}\sin\!\bigl(\delta k_x x_j+\delta k_z\hat{z}_j-\mu_r t+\psi\bigr)\hat{\sigma}^Z_j,
\end{multline}
where $\omega_s$ is the spin transition frequency, $\Omega$ is the microwave Rabi frequency, $\delta_{\mathrm{ac}}$ is the two-photon light shift arising from the interference of the two ODF beams, and $\psi$ is the relative phase between the ODF beams. The Pauli operators $\hat{\sigma}_j^X$ and $\hat{\sigma}_j^Z$ act on ion $j$, while the creation and annihilation operators $\hat{a}^\dagger_n$, $\hat{a}_n$ correspond to the axial mode $n$ with frequency $\omega_n$. Axial positions are expressed with the position operator $\hat{z}_j$. In contrast, the rotating ion position in the crystal plane is expressed classically in the laboratory frame as $x_j= r_j\cos(\omega_r t+\phi_j)$, with polar coordinates $(r_j,\phi_j)$ in the rotating frame. For our implementation, $\omega_r/2\pi = \qty{78}{\kilo\hertz}$ and $\Omega/2\pi = \qty{26}{\kilo\hertz}$.

In the microwave-dressed interaction frame and under a rotating-wave approximation, the combined drive of microwaves and ODF tuned to the rotational sideband at $\mu_r=\Omega+\omega_r = 2\pi \times \qty{104}{\kilo\hertz}$ produces an effective resonant coupling with a radius-dependent Rabi rate and an azimuth-dependent phase. This yields the effective Hamiltonian
\begin{equation}
\label{eqn:H_resonance}
\begin{aligned}
\tilde{H}_{\mathrm{init}}
=\frac{\delta_\mathrm{ac}\delta k_x}{8}\sum_j r_j &\Bigl((\hat{\sigma}_j^Z+\ i\hat{\sigma}_j^Y)e^{-i(\phi_j+\psi)}
\\&+(\hat{\sigma}_j^Z-i\hat{\sigma}_j^Y)e^{i(\phi_j+\psi)}\Bigr),
\end{aligned}
\end{equation}
As the coupling strength increases linearly with radius, we define the maximum Rabi rate attained at the crystal edge $r=R$ as
\begin{equation}
\label{eqn:site_dependent_Rabi_rate}
\Omega_{R}=\frac{\delta_\mathrm{ac}\,\delta k_xR}{2}.
\end{equation}
The Bloch vector of each ion $j$ then precesses at a local Rabi rate $\Omega_R\tilde{r}_j$, where $\tilde r_j=r_j/R$ is the normalised radius.

Equation~\eqref{eqn:H_resonance} therefore describes a sum of single-particle Hamiltonians. This non-interacting regime results from the present choice of ODF beatnote. By instead tuning $\mu_r$ to a sideband of the collective centre-of-mass mode, the spins can be coupled through their shared motion, enabling correlated dynamics and entanglement~\cite{Shankar2022}. Such interacting dynamics are not investigated in the present work. The single-spin Pauli operator expectation values for all spins initialised along $X$ evolve as
\begin{equation}
\begin{aligned}
\langle \hat{\sigma}_j^X(t)\rangle &= \cos(\tilde{r}_j\Omega_R t) \\
\langle \hat{\sigma}_j^Y(t)\rangle &= \sin(\tilde{r}_j\Omega_R t)\cos(\phi_j+\psi) \\
\langle \hat{\sigma}_j^Z(t)\rangle &= -\sin(\tilde{r}_j\Omega_R t)\sin(\phi_j+\psi).
\end{aligned}
\label{eqn:time_evolution_expectation_values}
\end{equation}

Figure \ref{fig1:experimental_implementation}c depicts the simulated spin evolution under Equation~\eqref{eqn:time_evolution_expectation_values}, showing that each spin undergoes a radius-dependent precession $\tilde r_j\Omega_R$. For a fixed $\psi$, the precession occurs about a local axis in the $Y-Z$ plane determined by the ion's azimuthal coordinate $\phi_j$. As $\tilde{H}_\mathrm{init}$ is applied, the radial gradient in precession speed and the azimuthally varying rotation axes cause the spins to diverge from $X$. This evolution develops a chiral texture that wraps progressively across the Bloch sphere. The target skyrmion texture is achieved at $\Omega_R t=\pi$, at which the peripheral spins ($\tilde r_j\simeq 1$) undergo a $\pi$-rotation and become anti-parallel to the central core, ensuring the spin texture fully wraps the sphere. For visualisation, we move to a rotated basis $(\mathcal{X},\mathcal{Y},\mathcal{Z})\equiv(Z,Y,-X)$ (Fig.~\ref{fig1:experimental_implementation}d).

This protocol can be modified by appending global microwave rotations and varying the application time of $\tilde{H}_\mathrm{init}$ to access a broader manifold of topological textures (see Appendix~\ref{additional_spin_textures}). For example, the system can produce fractional textures like merons as well as higher-order textures like the skyrmionium, consisting of a skyrmion nested within a second skyrmion of opposite topological charge. In addition, global rotations provide continuous control over the texture helicity and spin core, enabling the realisation of anti-skyrmions and bimerons \cite{Gobel2021, Shen2023}. 

\subsection*{Spin texture characterisation}\label{sec3}

To reconstruct the spin texture experimentally, we perform site-resolved projective measurements along the three principal Bloch sphere axes, as shown in Figure \ref{fig2:spin_texture_reconstruction}. Following the acquisition of a reference image taken after preparing all spins in $\ket{\uparrow}$, a gradient ascent algorithm is used to determine all ion positions. These coordinates are mapped onto the measurement image, and each ion's state is determined using time-resolved maximum likelihood estimation \cite{Wolk2015}. To account for shot-to-shot variations in the 2D crystal configuration, we apply polar binning to aggregate spin-up probabilities while preserving the underlying rotational symmetry of the texture. Local Bloch vectors are then reconstructed from these binned averages by performing three experimental projections along \{$X,Y,Z$\} to resolve the skyrmion texture. It is worth highlighting that this is not full state tomography as it omits inter-ion entanglement correlations. Instead, the skyrmion texture is characterised by the spatial configuration of the local Bloch vectors, allowing its topology to be determined directly from the resolved spin field. This approach is analogous to optical skyrmion measurements, where the topological winding is determined from the spatial variation of a local vector field~\cite{Tsesses2018,Deng2022}. 

We determine the required evolution time under $\tilde{H}_\mathrm{init}$ to obtain the skyrmion texture by evaluating the spin texture's topological winding number \cite{Fert2017},
\begin{equation}
\label{eqn:winding_number_integral}
Q = \frac{1}{4\pi}\int {\textbf{u}} \cdot \left(\frac{\partial {\textbf{u}}}{\partial x}\times\frac{\partial{\textbf{u}}}{\partial y}\right)\,dx dy,
\end{equation}
where ${\textbf{u}}(x,y)$ represents the local Bloch vector. $Q$ characterises the mapping between the spatial spin texture and a Bloch-sphere representation in which the Bloch vectors of all individual spins are represented on a common sphere. A measured value of $Q=\pm1$ indicates a non-trivial texture where the spin vectors span the entire surface of the Bloch sphere exactly once, consistent with the target skyrmion topology. This topological characterisation provides a more comprehensive assessment than measuring $\Omega_R$, as $Q$ captures the global spin texture by incorporating the configuration of all ions in the system. 

\begin{figure}[htbp]
    \includegraphics[width=\columnwidth]{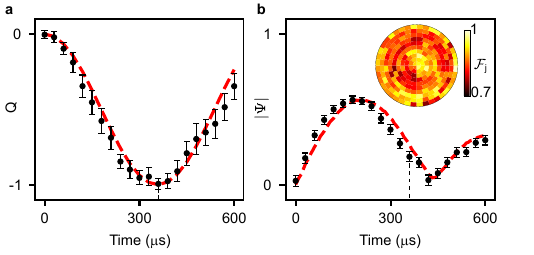}
    \caption{
    \textbf{Characterising the quality of the skyrmion spin texture.} Experimental (black dots) and theoretical (red dashed line) time evolution of (\textbf{a}) the topological winding number $Q$ and (\textbf{b}) the order parameter magnitude $|\Psi|$. The theory curve in (\textbf{a}) is fitted to extract $\Omega_R$, which is then used to predict the curve in (\textbf{b}). Points are calculated from the bin-averaged spin-projection expectation values ($n=200$ repetitions per projection). Error bars are $\pm1$ standard error, obtained by propagating the standard errors of the mean of the binned expectation values. The target skyrmion spin texture is obtained at $t=\qty{360}{\micro\second}$ when $Q=-1$ (vertical black dashed line). Inset: Average binned fidelity $\mathcal{F}_j$ of the experimental skyrmion texture at $t=\qty{360}{\micro\second}$ as a function of radius and azimuthal position. Radial and azimuthal positions are consistent with those in Fig. \ref{fig2:spin_texture_reconstruction}.}
    \label{fig3:spin_texture_fidelity}
\end{figure}

Experimentally, we track the evolution of the winding number as a function of Hamiltonian application time in Fig. \ref{fig3:spin_texture_fidelity}a, observing a minimum of $Q=-0.99 \pm 0.04$ at $t=\qty{360}{\micro\second}$. For the winding convention used here, a negative value signifies that the skyrmion's topological core is oriented downwards, whereas a positive winding would correspond to an upward orientation. While $Q$ is quantised to an integer when evaluated with a fixed spin boundary, the outermost spins in our system remain free to precess during the dynamics, allowing $Q$ to take non-integer values. From the measured winding-number dynamics $Q(t)$, we extract $\Omega_R$ by fitting to the theoretical spin trajectories calculated from Equation \eqref{eqn:time_evolution_expectation_values}, obtaining $\Omega_R/2\pi=1.56\pm \qty{0.07}{\kilo\hertz}$. This value sets the timescale of the spin-texture initialisation and corresponds to an ODF interaction strength $\delta_\mathrm{ac}$ consistent with previous measurements in our system (see Methods).

In addition to the topological winding number, we characterise the evolution of the spin texture using the time-dependent order parameter $\Psi(t)$. For a system of $N$ ions, this is defined as
\begin{equation}
\label{eqn:order_parameter}
\Psi(t) = \frac{1}{N}\sum_j{\tilde{r}_je^{i\phi_j} \langle \hat{\sigma}_j^Z-i\hat{\sigma}_j^Y\rangle}.
\end{equation}
A non-zero value indicates the development of a texture where the spin direction is correlated with the azimuthal angle of each ion. In the present work, the dynamics of $\Psi(t)$ reflect only the single-particle initialisation protocol, as no interactions are applied. However, during evolution under an interacting many-body Hamiltonian, the dynamics of $\Psi(t)$ can provide signatures of quantum correlations~\cite{Shankar2022}.

Using the spin trajectories from Equation \eqref{eqn:time_evolution_expectation_values} and previously obtained $\Omega_R$, we calculate the predicted evolution of the order parameter magnitude $|\Psi(t)|$ in Fig. \ref{fig3:spin_texture_fidelity}b. At the point of the skyrmion initialisation ($Q=-1$), we observe an experimental order parameter of $|\Psi|_\mathrm{exp}=0.29\pm 0.01$, showing good agreement with the theoretical value of $|\Psi|_\mathrm{th}=0.33$ (see Methods).

Lastly, we quantify the overall skyrmion initialisation and reconstruction accuracy by calculating the average local spin fidelity, $\bar{\mathcal{F}}$, which characterises the mean local Bloch vector overlap between the experimental spin texture and a pure-state target configuration \cite{Killoran2010},
\begin{equation}
\label{eqn:fidelity}
\bar{\mathcal{F}}=\frac{1}{N}\sum_j \mathcal{F}_j =\frac{1}{2N}\sum_j{\left(1+\textbf{u}_j\cdot\textbf{v}_j\right)},
\end{equation} 
where $\textbf{u}_j = (\langle\hat{\sigma}_j^X\rangle, \langle\hat{\sigma}_j^Y\rangle, \langle\hat{\sigma}_j^Z\rangle)$ and $\textbf{v}_j = (v_j^X, v_j^Y, v_j^Z)$ represent the reconstructed Bloch vector and the target Bloch vector for the $j$-th ion, respectively. 

Averaging across all polar bins, we obtain a fidelity of $\bar{\mathcal{F}}=0.87\pm 0.03$ for the spin texture with $Q\approx-1$. Errors in state estimation contribute approximately $2\%$ to the total error, determined by calculating the average infidelity of reference images taken of the crystal in fully bright and dark configurations \cite{Wolf2024}. An additional $2\%$ error arises from Hamiltonian approximations that omit higher-order terms present in the experimental drive (see Appendix~\ref{sec_Hamiltonian_derivation}). We attribute the remaining error to off-resonant scattering, dephasing from finite in-plane motional temperature, and inhomogeneities in the ODF intensity.

\begin{figure*}[!htbp]
    \centering
    \includegraphics[width=\textwidth]{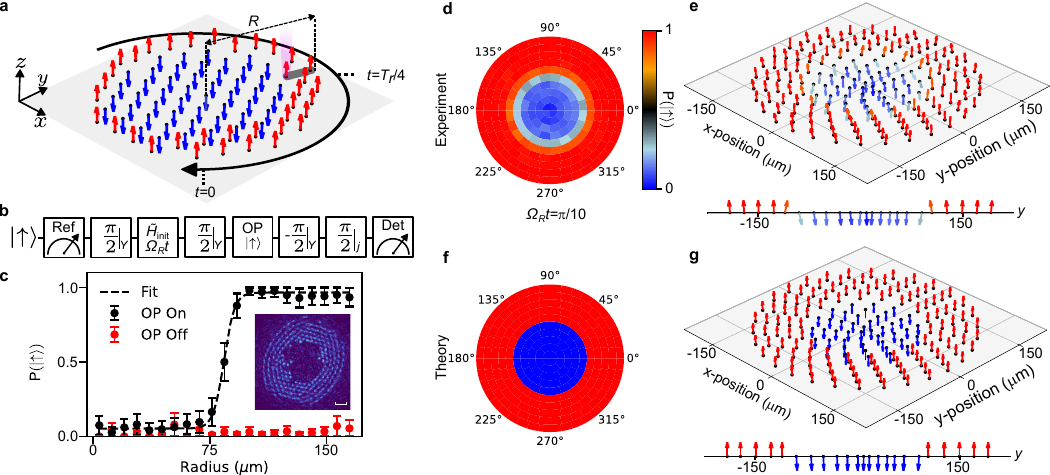}
  \captionof{figure}{\textbf{Initialisation of a domain wall.} %
  (\textbf{a}) Schematic for the implementation of the domain wall. A focused laser with a beam waist on the order of the ion-ion spacing ($\sim\qty{30}{\micro\meter}$) is used to selectively perform optical repumping. The natural rotation of the ion crystal causes the beam to interact with all ions at a targeted radius over a rotation period of the crystal $T_r=2\pi / \omega_r$. An acousto-optic modulator is used to move the beam radially along the crystal to target ions at a different radius. %
  (\textbf{b}) Pulse sequence for the domain wall initialisation. Here, $\tilde{H}_\mathrm{init}$ is driven for a duration of $\Omega_R t=\pi /10$ to initialise a small order parameter. A $\pi/2|_Y$ pulse is used to rotate the texture to the regular spin basis and selective optical pumping (OP) is then applied to the outer ions with $r_j\geq R/2$. Next, a $-\pi/2|_Y$ pulse rotates all spins back to the rotated basis before measurement. %
  (\textbf{c}) Spin state probability $P(\ket{\uparrow})$ as a function of ion radius after selective optical pumping (black) and without (red). Each point is the mean $P(\ket{\uparrow})$ of all ions within a radial bin, averaged over $n=200$ experimental repetitions. Error bars are $\pm1$ standard error of the mean over the repetitions. The dashed black line is the fit to a sigmoid error function. %
  (\textbf{d}) Experimental probability distribution of the $\mathcal{Z}$-projection domain wall, and (\textbf{e}) the corresponding reconstructed spin texture. %
   (\textbf{f}) and (\textbf{g}) show the theoretical $\mathcal{Z}$-projection and reconstructed spin texture, respectively.
   }
  \label{fig4:domain_wall_characterisation}
\end{figure*}

\subsection*{Domain wall initialisation}\label{sec4}

Beyond skyrmion textures, we demonstrate the deterministic preparation of Ising-type domain walls, consisting of two regions of opposite spin polarisation separated by a sharp boundary. As illustrated in Fig.~\ref{fig4:domain_wall_characterisation}a, we initialise a domain wall at radius $r=R/2$ using a two-step protocol. We first apply $\tilde{H}_\mathrm{init}$ for a short duration to generate a small, non-zero order parameter. We then use a focused optical pumping beam to selectively reset ions at larger radii to $\ket{\uparrow}$.

The full sequence, shown in Fig.~\ref{fig4:domain_wall_characterisation}b, extends the skyrmion preparation protocol by incorporating spatially resolved local control via a tightly focused optical pumping beam steered with an acousto-optic modulator. For the larger $220$-ion crystal used in this experiment, the crystal radius is $R\simeq\qty{170}{\micro\meter}$, and we initialise the spin boundary at $R/2\simeq\qty{85}{\micro\meter}$. The beam is swept radially inward from $r\simeq\qty{170}{\micro\meter}$ to $\qty{85}{\micro\meter}$, with a pulse applied at each position for four crystal rotation periods ($\approx\qty{51.2}{\micro\second}$). This ensures uniform exposure of ions along each addressed radius (see Methods).

We characterise the resulting boundary by measuring the spin-up probability $P(\ket{\uparrow})$ as a function of radius, as shown in Fig.~\ref{fig4:domain_wall_characterisation}c. The $10$--$90\%$ edge width is $28\pm\qty{12}{\micro\meter}$, comparable to the inter-ion spacing. Residual population near the domain boundary arises primarily from the finite beam waist, which partially addresses neighbouring ions. Deviations from perfect circular symmetry of the crystal result in additional broadening, leading to non-uniform beam exposure at the domain boundary.

To quantify the prepared state, we reconstruct the local Bloch vectors using the protocol described above. We obtain an order parameter of $|\Psi|_\mathrm{exp}=0.07\pm0.02$ and a mean state fidelity of $\bar{\mathcal{F}}=0.93\pm0.02$. The measured order parameter is in agreement with the theoretical value of $|\Psi|_\mathrm{th}=0.04$. As shown in Figs.~\ref{fig4:domain_wall_characterisation}d,f, the domain boundary is the primary source of fidelity loss, accounting for $3\%$ of the total error. This is consistent with the finite beam waist and geometric effects discussed above. The remaining infidelity is attributed to errors in state estimation. Figs.~\ref{fig4:domain_wall_characterisation}e,g show the reconstructed domain wall and vertical cross section around $x=0$, which both display good overall agreement with theory. This capability establishes local control within large ion crystals and enables the controlled preparation of spatially structured initial states for studying nonequilibrium many-body dynamics.

\section*{Discussion}\label{sec5}

We have demonstrated deterministic preparation and site-resolved reconstruction of spatially structured spin textures in a two-dimensional Penning-trap ion crystal, including both skyrmion and domain-wall configurations. This is enabled by combining wavefront-engineered spin-dependent forces, spatially resolved local control, and high-resolution imaging within a large ion crystal. These results establish the Penning-trap platform as a system capable of accessing spatially structured spin states beyond the permutation-symmetric regime. While the present work focuses on initialising product states, $\tilde{H}_\mathrm{init}$ can more generally serve as a programmable resource in a variational quantum circuit. Interleaving $\tilde{H}_\mathrm{init}$ with global single-qubit rotations and established entangling operations such as light-shift gates~\cite{Britton2012, Bohnet2016} or a M\o lmer-S\o rensen interaction~\cite{Carter2023} could enable preparation of spatially structured correlated many-body states beyond the permutation-symmetric regime and provide a route to studying two-dimensional nonequilibrium dynamics in interacting quantum systems. 

This level of control enables direct experimental investigation of how topology, winding number, and spatial order influence far-from-equilibrium dynamics in interacting quantum systems. More broadly, site-resolved measurements in large programmable simulators provide a route to identifying emergent phases and their characteristic correlations directly from experimental snapshots, such as those in Fig. \ref{fig1:experimental_implementation}b. Recent work has shown that such approaches can reveal hidden order parameters, unconventional critical behaviour, and distinct dynamical regimes through correlation-based analysis and interpretable machine-learning techniques \cite{Chiu2019, Rispoli2019, Greplova2020, Miles2023}.

A natural application of this platform is the study of dynamical phases in chiral $p$-wave systems \cite{Shankar2022}. Up to a $\pi$-pulse, the initial states demonstrated here directly realise the configurations required for these systems: skyrmion states enable access to phase I and II dynamics, while domain-wall states are required for phase III. In this mapping, the initial spin configuration encodes the momentum-space structure of paired states, while subsequent evolution is generated by driving near the centre-of-mass mode. The detuning from this mode acts as a tunable quench parameter, and the control fields and the collective mode itself provide an effective bosonic channel \cite{Shankar2022}.

Extending these studies to longer evolution times will require suppression of decoherence from off-resonant scattering and low-frequency magnetic-field fluctuations. While off-resonant scattering is small during state preparation, it limits coherence during dynamical evolution and must be reduced to evolve the system over appreciable times. This can be achieved through optimisation of the ODF operating point \cite{Carter2023} and increased beam opening angles $\theta_\mathrm{ODF}$ \cite{Pham2024}. In addition, cooling of radial modes \cite{Johnson2024, Johnson2025} would reduce dephasing due to finite motional temperature, preserving spin contrast during both initialisation and subsequent dynamics.

Looking ahead, full single-ion addressing offers a complementary approach to preparing topological spin textures~\cite{Polloreno2022, Carter2026}. In our system, this has been realised by adapting the focused optical pumping beam to induce controlled AC Stark shifts synchronised with the crystal rotation~\cite{Makadia2026}. Combined with global microwave operations, this enables deterministic local spin control with minimal crosstalk. Beyond state preparation, single-ion addressing would also be essential for future readout schemes based on spatially resolved spin cross-correlations. Together, single-ion addressing and ODF-based control offer a route to preparing and probing topological spin textures by combining spatial control, collective interactions, and site-resolved readout in a single many-body quantum simulator.

\section*{Methods}

\subsection*{ODF phase stabilisation}
\label{sec_odf_phase_stabilisation}

To ensure shot-to-shot reproducibility of the spin textures, the relative phase between the $\mathcal{X}$- and $\mathcal{Y}$-projection dipole patterns must remain stable. This dipole orientation is determined by the relative phase $\psi$ of the two ODF beams. However, environmental fluctuations, such as changes in temperature and air density between the interferometer arms, cause variations in optical path lengths, leading to phase instability of the optical interference pattern.

We stabilise $\psi$ using an optical phase-locked loop. A radiofrequency (RF) reference beat note is generated by coupling out and mixing a small part of the two RF signals used to drive the ODF acousto-optical modulators. Similarly, a fraction of the ODF beams is picked off before entering the magnet bore and interfered on a photodiode, producing a corresponding optical beat note signal. Mixing the RF and optical beat note signals yields an error signal proportional to the phase difference between them. This error signal is processed by a proportional-integral-derivative (PID) feedback controller, whose output drives a piezoelectric-actuated mirror in one of the ODF arms. Feedback to the mirror adjusts the optical path length to stabilise the optical beat-note phase relative to the RF reference. Along the remaining beam path in the magnet bore, both ODF beams travel in close proximity along approximately the same path, thereby making fluctuations in the bore predominantly common-mode.

Although $\psi$ is actively stabilised, its absolute reference with respect to the laboratory coordinate frame is not known a priori. A change in this reference frame is equivalent to a global rotation of the $\mathcal{X}$- and $\mathcal{Y}$-projection dipole patterns, and therefore appears as an offset in the ODF phase when expressed in the laboratory frame. The desired alignment can be recovered either experimentally by adjusting the relative phase between the ODF beams, or in post-processing by applying a common azimuthal offset $\varphi$ to all ion coordinates $\phi_j\rightarrow\phi_j+\varphi$. We adopt the latter approach and determine the optimal offset by maximising the texture fidelity with respect to the target skyrmion state,
\begin{equation}
    \varphi_{\mathrm{opt}}
    = \underset{\varphi}{\arg \operatorname{max}}\,
    \bar{\mathcal{F}}(\varphi).
\end{equation}
This procedure optimises only the global orientation of the reconstructed texture and leaves $Q$ and $|\Psi(t)|$ invariant.

\subsection*{Winding number calculation}\label{sec_winding_number_calc}

\begin{figure}[htbp]
    \centering
  \includegraphics[width=\columnwidth]{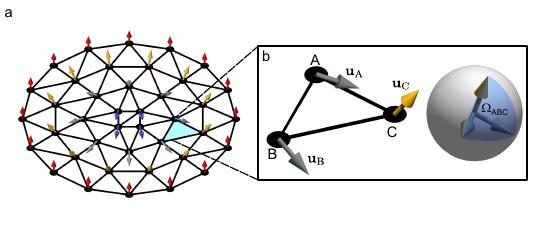}
  \caption{\textbf{Delaunay triangulation for winding number calculation.} %
  (\textbf{a}) A triangular mesh is constructed using Delaunay triangulation, forming a non-overlapping network with spins located at the vertices. Arrows indicate the spin orientation as in Fig.~\ref{fig2:spin_texture_reconstruction}. (\textbf{b}) Enlarged view of the shaded simplex in (\textbf{a}), spanned by vertices $\mathrm{A},\mathrm{B}$, and $\mathrm{C}$. The associated unit vectors $\textbf{u}_\mathrm{A}$, $\textbf{u}_\mathrm{B}$, and $\textbf{u}_\mathrm{C}$ define the vertices of a spherical triangle. The solid angle $\Omega_\mathrm{ABC}$ is equal to the area subtended by this spherical triangle on the unit sphere~\cite{Oosterom1983}.
  }
  \label{fig5:delaunay_triangulation}
\end{figure}

The winding number, or topological charge $Q$, is typically defined as a surface integral over the spin texture, as given in Equation \eqref{eqn:winding_number_integral}. However, as the ion crystal samples the spin field only at discretised positions, we instead evaluate $Q$ using Delaunay triangulation \cite{Yin2016} (see Fig.~\ref{fig5:delaunay_triangulation}). The triangulation forms a non-overlapping mesh whose vertices are neighbouring ions. For each triangle $\triangle_\mathrm{ABC}$, the three local spin unit vectors $\textbf{u}_\mathrm{A}$, $\textbf{u}_\mathrm{B}$, and $\textbf{u}_\mathrm{C}$ define a spherical triangle on the Bloch sphere, with solid angle
\begin{equation}
\label{eqn:Delaunay_angle}
\Omega_\mathrm{ABC}= 2\tan^{-1}\left(\frac{\textbf{u}_\mathrm{A}\cdot(\textbf{u}_{\mathrm{B}}\times\textbf{u}_\mathrm{C})}{1+\textbf{u}_\mathrm{A}\cdot\textbf{u}_\mathrm{B}+\textbf{u}_\mathrm{B}\cdot\textbf{u}_\mathrm{C}+\textbf{u}_\mathrm{C}\cdot\textbf{u}_\mathrm{A}}\right).
\end{equation}
The winding number is then obtained by summing $\Omega_\mathrm{ABC}$ over all triangles,
\begin{equation}
\label{eqn:winding_number_discrete}
Q = \frac{1}{4\pi}\sum_{\triangle_\mathrm{ABC}}{\Omega_\mathrm{ABC}}.
\end{equation}
The orientation of the skyrmion texture determines the sign of the winding number. A positive (negative) value of $Q$ corresponds to a skyrmion state where the central spins align with $\ket{\uparrow}_\mathcal{Z}$ ($\ket{\downarrow}_\mathcal{Z}$) and gradually tilt towards $\ket{\downarrow}_\mathcal{Z}$ ($\ket{\uparrow}_\mathcal{Z}$) with increasing crystal radius.

When calculating the experimental winding number, the measured Bloch vectors are first normalised before evaluating the solid angles in Equation~\eqref{eqn:winding_number_discrete}. Experimentally, measured Bloch vectors have reduced magnitude due to statistical uncertainty arising from the finite number of measurement shots. Because the winding number depends only on the orientation of the local spin field, we normalise the measured Bloch vectors prior to evaluating $Q$, provided their magnitudes are non-zero. Loss of Bloch vector magnitude is quantified separately through the fidelity.

To compare the experimentally measured winding-number dynamics with the theoretical model, we evaluate the spin texture evolution using the trajectories given in Equation~\eqref{eqn:time_evolution_expectation_values}. The spin trajectories depend on the spatial coordinates of the ions $\left(r_j,\phi_j\right)$, and the maximum Rabi frequency $\Omega_R$, which sets the timescale of the spin rotation under $\tilde{H}_\mathrm{init}$. We use experimentally measured ion positions from a crystal of $N=160$ ions and calculate the corresponding winding number using the same Delaunay-triangulation procedure applied to the experimental data.

From Equation~\eqref{eqn:site_dependent_Rabi_rate}, $\Omega_R$ is set by the crystal radius $R$, the in-plane component of the ODF difference wavevector $\delta k_x$, and the ODF potential amplitude $\delta_\mathrm{ac}$. While $R$ and $\delta k_x$ can be readily measured, $\delta_\mathrm{ac}$ is difficult to determine due to the limited access to the vacuum chamber. We therefore determine $\Omega_R$ by fitting the experimentally measured winding-number evolution to the theoretical model, and use the fitted value to infer $\delta_\mathrm{ac}=2\pi\times\qty{4.8}{\kilo\hertz}$. 

Combining this value with the measured Debye--Waller factor $\mathrm{DWF}\approx0.8$, we obtain an ODF magnitude of $F_0\approx\qty{16}{\yocto\newton}$, where
\begin{equation}
\label{eqn:F0_equation}
F_0 = \hbar\delta_\mathrm{ac}|\delta \textbf{k}| \mathrm{DWF}. 
\end{equation}
This value is consistent with previous measurements of the ODF in our system~\cite{Pham2024}, providing an independent validation of the drive strength fitted from the winding number dynamics.

\subsection*{Focused optical pumping beam}
\label{sec_SIR_beam}

\begin{figure}[!htbp]
  \includegraphics[width=\columnwidth]{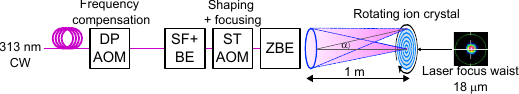}
  \centering
  \caption{\textbf{Focused optical pumping beam schematic.} \qty{313}{\nano\meter} continuous-wave (CW) light is coupled from a fibre into a double-pass acousto-optic modulator (DP AOM) for frequency compensation. After spatial filtering and beam expansion (SF + BE), a single-pass steering AOM (ST AOM) deflects the beam by angle $\alpha$ radially across the ion crystal. A zoom beam expander (ZBE) adjusts the beam waist before the beam is propagated over a \qty{1}{\meter} path length and focused to a waist of \qty{18}{\micro\meter} at the ion crystal.
  }
  \label{fig6:SIR_schematic}
\end{figure}

We implement a focused repump beam to enable local spin control within the ion crystal (see Figure \ref{fig6:SIR_schematic}). The beam can be steered radially across the crystal, allowing selective addressing of ions over a range extending to $r\approx\qty{250}{\micro\meter}$.

Control of the beam is achieved using two acousto-optic modulators (AOMs). The beam first passes through a double-pass ``compensation'' AOM, followed by a single-pass ``steering'' AOM. The steering AOM deflects the beam radially, enabling selective addressing of ions at different radii. As this deflection introduces frequency and power variations, the double-pass AOM provides a compensating frequency offset and stabilises the optical power, maintaining consistent beam parameters across the crystal.

The beam is aligned to the crystal centre using a piezoelectric mirror mount, with the position monitored via the TPX3CAM. During operation, the optical power is dynamically adjusted: higher power is used at larger radii to broaden the effective addressed region and repump multiple ion rings simultaneously, while reduced power near the domain boundary minimises unintended excitation beyond the target region.

This approach is related to the scheme demonstrated by McMahon \textit{et al.}~\cite{McMahon2024}; here, however, the beam is repositionable by an AOM and tuned to drive a repump transition rather than induce an AC Stark shift.

\subsection*{Theoretical fitting of the order parameter}
\label{theoretical_fitting}

We derive the theoretical behaviour of the order parameter by numerically simulating the single-particle dynamics under the microwave-dressed interaction-frame Hamiltonian (Equation~\ref{eqn:H_dressed}), obtained from Equation~\eqref{eqn:H_init}, after neglecting coupling to the far-detuned axial motional modes. As the Hamiltonian consists only of single-particle interactions, the dynamics can be simulated efficiently using classical hardware. We perform simulations using the parameters given in Table~\ref{tab:skyrmion_experimental_parameters} for a 160-ion crystal, utilising spatial positions taken directly from experimentally measured ion coordinates.

\begin{table}[htbp]
  \centering
  \caption{\textbf{Experimental parameters used in the theoretical order parameter simulations.} $\delta_\mathrm{ac}$ is estimated from a fit to the measured winding-number evolution.}
  \label{tab:skyrmion_experimental_parameters}
  \begin{tabularx}{\columnwidth}{c >{\raggedright\arraybackslash}X c}
    \hline\hline
    \textbf{Parameter} & \textbf{Description} & \textbf{Value} \\
    \midrule
    $\Omega$ & Microwave Rabi rate & $2\pi\times\qty{26}{\kilo\hertz}$ \\
    $\omega_r$ & Crystal rotation frequency & $2\pi\times\qty{78}{\kilo\hertz}$ \\
    $\mu_r$ & ODF beatnote frequency & $2\pi\times\qty{104}{\kilo\hertz}$ \\
    $\delta_\mathrm{ac}$ & ODF two-photon light shift & $2\pi\times\qty{4.8}{\kilo\hertz}$ \\
    $\psi$ & Relative phase between the ODF beams & $\qty{90}{\degree}$ \\
    $\theta_\mathrm{ODF}$ & ODF beam crossing angle & $\qty{18}{\degree}$ \\
    $\delta\theta$ & Tilt of $\delta\textbf{k}$ from the crystal-plane normal & $\qty{0.04}{\degree}$ \\
    $R$ & Crystal radius & $\qty{150}{\micro\meter}$ \\
    \hline\hline
  \end{tabularx}
\end{table}

We prepare the initial spin state of all ions to point along $+X$. The individual spin states are then evolved under the single-particle Hamiltonian $H_\mathrm{init}$. The time-evolution expectation values of $\{\hat\sigma_j^X,\hat\sigma_j^Y,\hat\sigma_j^Z\}$ for each ion $j$ are determined, corresponding to the three experimentally measured spin projections. To model the finite-shot nature of the experiment, we simulate measurement outcomes of $\{-1,1\}$ through Monte Carlo sampling. For each ion and each spin projection, 200 independent samples are drawn according to the probability distribution $P(\pm1)=\frac{1\pm\langle\hat\sigma_j^i\rangle}{2}$, for $i\in\{X,Y,Z\}$. No correlations are imposed between projections, such that each projection represents an independently measured shot.

Following the generation of the simulated measurements, the full crystal is partitioned into polar bins using the same binning procedure described in the main text, where the crystal is divided into 10 equal-width radial annuli, with each annulus further subdivided into angular sub-bins, totalling 220 bins. Empty bins are excluded from analysis. Within each bin, spin projection measurements are averaged to obtain the local binned expectation values. The order parameter is then calculated from the resulting spin textures using the bin positions and averaged spin projections according to Equation~\eqref{eqn:order_parameter}.

The above procedure is repeated over multiple independent simulated experimental runs. To account for the absence of a fixed crystal orientation relative to the laboratory ODF phase at the start of each experimental repetition, we randomise the global azimuthal offset of the crystal by applying $\phi_j\rightarrow\phi_j+\phi_\mathrm{offset}$, where the offset $\phi_\mathrm{offset}$ is sampled uniformly from the interval $[0,2\pi)$. The rotated ion coordinates are used to construct the Hamiltonian, while the polar-bin coordinates remain fixed. This reproduces the shot-to-shot variation in the crystal orientation present experimentally without averaging away the texture defined in the co-rotating crystal frame. The theoretical curves shown in Figure~\ref{fig3:spin_texture_fidelity} are obtained by averaging the calculated order parameter over 20 simulated realisations.

\begin{figure}[!htbp]
    \centering
    \includegraphics[width=\columnwidth]{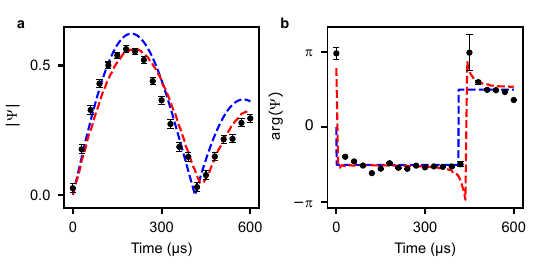}
  \captionof{figure}{\textbf{Comparison of theoretical models for the order parameter.} (\textbf{a}) Magnitude $|\Psi|$ and (\textbf{b}) phase $\arg{(\Psi)}$ of the order parameter as a function of evolution time. Black points show the experimental values calculated from the bin-averaged spin-projection expectation values ($n=200$ repetitions per projection). Error bars are $\pm1$ standard error, obtained by propagating the standard errors of the mean of the binned expectation values. The red dashed curves show the Monte Carlo simulation under the microwave-dressed-frame Hamiltonian given in Equation~\ref{eqn:H_dressed}. The magnitude curve in (\textbf{a}) is reproduced from Fig.~\ref{fig3:spin_texture_fidelity}b. The blue dashed curves show the corresponding prediction obtained from the approximate analytic model after the small-angle and rotating-wave approximations. 
   }
  \label{fig7:order_parameter_modelling}
\end{figure}

Figure~\ref{fig7:order_parameter_modelling} compares the resulting dynamics from the Monte Carlo simulation of the microwave-dressed-frame Hamiltonian with both the experimental data and the approximate spin trajectories from Equation~\eqref{eqn:time_evolution_expectation_values}. The phase $\arg(\Psi)$ is determined after applying the same global azimuthal phase convention used to align the reconstructed texture. The dressed-frame simulation reproduces the experimentally observed evolution of both the magnitude $|\Psi|$ and phase $\arg(\Psi)$ more closely than the approximate analytic model, including the observed peak amplitudes in $|\Psi|$, and the phase jump in $\arg(\Psi)$.

\section*{Data Availability}
The experimental data presented in this article are available via the Zenodo repository~\cite{Jee2026Data} (\href{https://doi.org/10.5281/zenodo.22804849}{doi:10.5281/zenodo.22804849}).

\section*{Acknowledgements}
We would like to thank Yasir Iqbal, Cameron McGarry, and John Bollinger for valuable feedback on the manuscript. 

\section*{Funding Statement}
This material is based upon work supported by the Air Force Office of Scientific Research (FA2386-23-1-4067), the Australian Research Council Centre of Excellence for Engineered Quantum Systems (CE170100009) and a private grant from H. and A. Harley. R.N.W. acknowledges support from the Australian Research Council under the Discovery Early Career Researcher Award scheme (DE190101137). We acknowledge support from the Sydney Quantum Academy (J.Y.Z.J., J.H.P., and G.C.M.) and the Australian Government Research Training Program (RTP) Scholarship (J.Y.Z.J., J.H.P., N.M.). A.S. acknowledges support by the Department of Science and Technology, Govt. of India through the INSPIRE Faculty Award (DST/INSPIRE/04/2023/001486), by the Anusandhan National Research Foundation (ANRF), Govt. of India through the Prime Minister’s Early Career Research Grant (PMECRG) (ANRF/ECRG/2024/001160/PMS) and by IIT Madras through the New Faculty Initiation Grant (NFIG).

\section*{Author Contributions}
The experiment was performed by J.Y.Z.J., N.M., G.C.M., J.H.P., and R.N.W. J.Y.Z.J. analysed the data. A.S. provided the theory. J.Y.Z.J. and R.N.W. wrote the manuscript with feedback from N.M., J.H.P., G.C.M., M.J.B., and A.S. R.N.W. supervised the project.

\section*{Competing Interests}
The authors declare no competing interests.

\appendix
\renewcommand{\thesection}{\Alph{section}}

\titleformat{\section}[display]
  {\normalfont\normalsize\bfseries\centering}
  {Appendix \thesection:}{0.5em}{}

\section{Derivation of the initialisation Hamiltonian}
\label{sec_Hamiltonian_derivation}
To derive the effective Hamiltonian $\tilde{H}_\mathrm{init}$, we begin with Equation 1 from the main text and move to the microwave-dressed interaction frame, 
\begin{equation}
\label{eqn:H_dressed}
H_{\mathrm{init}} = \sum_j{\delta_\mathrm{ac}\sin(\delta k_xx_j-\mu_rt+\psi)\\(\tilde{\hat{\sigma}}_j^+e^{-i\Omega t}+\tilde{\hat{\sigma}}_j^-e^{i\Omega t})},
\end{equation} 
with operators $\tilde{\hat{\sigma}}_j^\pm=\frac{1}{2}(\hat{\sigma}_j^Z\pm i\hat{\sigma}_j^Y)$. The axial-mode frequencies can be neglected as the ODF beams are far detuned from the modes in this work. Assuming $\delta k_x R\ll1$, we expand the Hamiltonian using the approximation $\sin(A-B)\approx A\cos(B)-\sin(B)$, where $A=\delta k_x x_j$ and $B=\mu_rt-\psi$. The second term can be neglected as it is rapidly oscillating. In our experiment, $\delta k_x R\simeq0.66$. Numerically comparing evolution under the exact and small-angle Hamiltonians for $\delta k_x R\simeq0.66$ gives an approximately 2\% reduction in state fidelity. While reducing $\delta k_x$ would further validate this approximation, it would necessitate a longer evolution time, thereby increasing the system's susceptibility to decoherence and worsening the overall infidelity.

Next, writing the ion position in polar coordinates $x_j=r_j\cos(\omega_rt+\phi_j)$ and expanding $\cos(B)$ in terms of exponentials, we obtain the following Hamiltonian,

\begin{equation}
\label{eqn:H_expanded}
\begin{aligned}
\tilde{H}_{\mathrm{init}}
= \sum_j \frac{\delta_\mathrm{ac}\delta k_x r_j}{4}
\Bigl(&
\tilde{\hat{\sigma}}_j^+
e^{i[(\omega_r+\mu_r-\Omega)t+(\phi_j-\psi)]}
\\
&+
\tilde{\hat{\sigma}}_j^+
e^{i[(\omega_r-\mu_r-\Omega)t+(\phi_j+\psi)]}
\\
&+
\tilde{\hat{\sigma}}_j^+
e^{-i[(\omega_r-\mu_r+\Omega)t+(\phi_j+\psi)]}
\\
&+
\tilde{\hat{\sigma}}_j^+
e^{-i[(\omega_r+\mu_r+\Omega)t+(\phi_j-\psi)]}
\Bigr)
+\mathrm{h.c.}
\end{aligned}
\end{equation}
By choosing $\mu_r=\Omega+\omega_r$, the third term is left resonant, and all others become off-resonant. For the other terms to be considered negligible, the maximum Rabi rate $\Omega_R$ must be small compared to the off-resonant frequencies such that the rotating-wave approximation holds,
\begin{equation}
\label{eqn:rwa_approx}
\Omega_R \ll |2\omega_r|, |2\Omega|, |2(\omega_r+\Omega)|. 
\end{equation}
For our experimental parameters, we set $\Omega_R/2\pi=\qty{1.56}{\kilo\hertz}$, $\omega_r/2\pi=\qty{78}{\kilo\hertz}$, and $\Omega/2\pi=\qty{26}{\kilo\hertz}$. Comparing these values, $|2\Omega|$ represents the lowest frequency threshold in the rotating-wave approximation condition ($\Omega_R/2\pi\ll\qty{50}{\kilo\hertz}$). As $\Omega_R$ is an order of magnitude smaller than this limit, the off-resonant contributions are negligible. 

Rewriting Equation \eqref{eqn:H_expanded} with only resonant terms, we can simplify $\tilde{H}_{\mathrm{init}}$ to
\begin{equation}
\label{eqn:H_resonance_2}
\tilde{H}_{\mathrm{init}} = \sum_{j}\frac{\Omega_R\tilde{r}_j}{2}(\tilde{\hat{\sigma}}_j^+e^{-i(\phi_j+\psi)}+\tilde{\hat{\sigma}}_j^-e^{i(\phi_j+\psi)}),
\end{equation}
expressed in terms of normalised radii $\tilde{r}_j$. Next, denoting 
\begin{equation}
\label{eqn:expectation_values}
\textbf{v}_j=\left(\langle\hat{\sigma}_j^X\rangle,\langle\hat{\sigma}_j^Y\rangle,\langle\hat{\sigma}_j^Z\rangle\right)^T,
\end{equation}
the equations of motion for the spin expectation values are given by
\begin{equation}
\label{eqn:equations_of_motion}
\frac{d\textbf{v}_j}{dt}=M_j\textbf{v}_j
\end{equation}
where the matrix $M_j$ is given by
\begin{equation}
\label{eqn:H_init_matrix}
M_j=\Omega_R\tilde{r}_j\begin{pmatrix}
0 & -\cos(\chi_j) & \sin(\chi_j)\\
\cos(\chi_j) & 0 & 0\\
-\sin(\chi_j) & 0 & 0
\end{pmatrix},
\end{equation}
and $\chi_j=\phi_j+\psi$.
These equations constitute a set of three coupled first-order differential equations that can be solved to give the expectation values in Equation 4 of the main text.

\section{Programmable spin textures}
\label{additional_spin_textures}

\begin{figure}[!thp]
    \centering
  \includegraphics[width=\columnwidth]{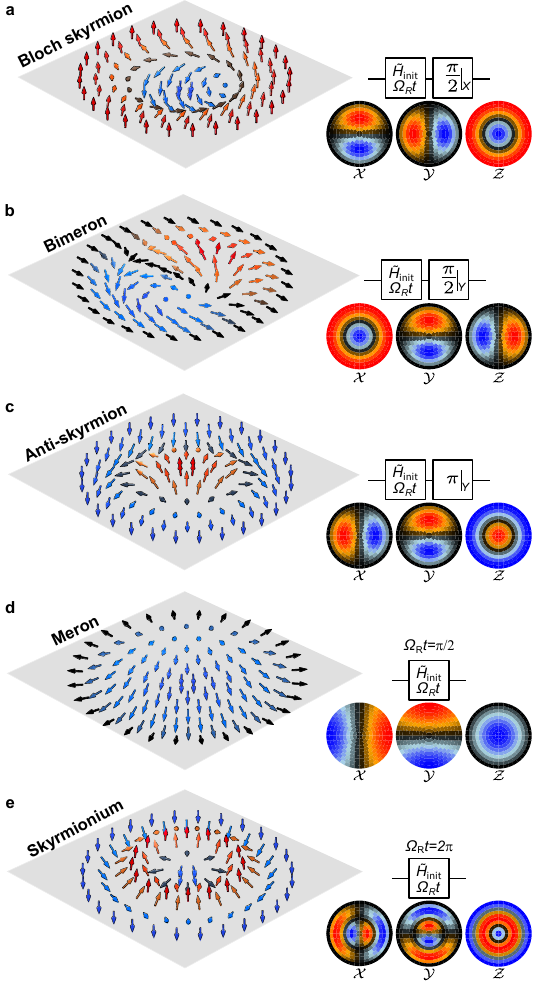}
  \caption{\textbf{Programmable initialisations of various topological spin textures.} %
   Theoretical plots of various spin textures that can be reached by modifying the skyrmion initialisation protocol, with the pulse-sequence modification and state projections in the rotated $(\mathcal{X},\mathcal{Y},\mathcal{Z})$ basis shown alongside. %
   (\textbf{a}-\textbf{c}) Spin textures with $Q=-1$, initialised from the N\'eel skyrmion via global spin rotations. Specifically, (\textbf{a}) a Bloch skyrmion is realised by a $\pm\pi/2|_X$ rotation; the choice of rotation determines the helicity $\gamma$. Similarly, the (\textbf{b}) bimeron and (\textbf{c}) anti-skyrmion textures are obtained by applying a $\pi/2|_Y$ and a $\pi|_Y$ pulse, respectively. %
   (\textbf{d}-\textbf{e}) Additional spin textures generated by modulating the duration of the initialisation drive. The (\textbf{d}) meron and (\textbf{e}) skyrmionium are obtained by driving $\tilde{H}_\mathrm{init}$ for $t=\pi/(2\Omega_R)$ and $t=2\pi/\Omega_R$, respectively.
  }
  \label{supp_fig1:additional_spin_textures}
\end{figure}

The controlled Hamiltonian $\tilde{H}_\mathrm{init}$ enables preparation of a broader family of topological textures beyond the N\'eel-type skyrmion \cite{Nagaosa2013} with minor modifications to the experimental protocol outlined in the main text. In Figure \ref{supp_fig1:additional_spin_textures} we show theoretical topological spin textures achieved through varying the parameters of the controlled initialisation interaction.

The Bloch-type skyrmion, bimeron, and anti-skyrmion (Figure \ref{supp_fig1:additional_spin_textures}a-c) are related to the N\'eel skyrmion via Bloch basis transformations \cite{Kharkov2017, Psaroudaki2021, Goerzen2025}, and their configurations are accessible via global spin rotations after the initial driving sequence. A global $R_x$ rotation enables continuous tuning of the texture's helicity, $\gamma$, which is the internal phase offset between the spatial azimuthal angle and the spin orientation. Varying this rotation smoothly interpolates between the N\'eel-type configurations ($\gamma=0, \pi$), where spins wind radially, and Bloch-type configurations ($\gamma=\pm\pi/2$), where spins wind tangentially.

Rotations about $\mathcal{X}$ and $\mathcal{Y}$ further control the texture's polarity, the direction of the spin core relative to the crystal plane. A global $\pi/2|_Y$ `tilts' the skyrmion into the plane to form a bimeron. This transformation is equivalent to viewing the skyrmion in the regular Bloch basis $\left(X,Y,Z\right)$, mapping it into a dipolar in-plane structure. Additionally, a $\pi|_Y$ rotation maps the skyrmion to an anti-skyrmion by inverting its polarity. Importantly, the topological charge is preserved under a global rotation, retaining $Q=-1$. 

In addition to basis transformations, $Q$ can be controlled by varying the drive duration $t$ of $\tilde{H}_\mathrm{init}$, which governs the rotation of the spin texture's polar angle. A drive duration of $t=\pi/\left(2\Omega_R\right)$ generates a meron (Figure \ref{supp_fig1:additional_spin_textures}d) with a fractional topological winding number of $Q=-1/2$, while extending the drive to $t=2\pi/\Omega_R$ generates higher-order textures such as the skyrmionium (Figure \ref{supp_fig1:additional_spin_textures}e). This configuration consists of a skyrmion nested within a second skyrmion of opposite polarity \cite{Zhang2018}. The resulting texture is topologically trivial ($Q=0$), yet retains a non-trivial local spin configuration.

In conventional systems, these textures demand carefully engineered magnetic anisotropies or specific Dzyaloshinskii-Moriya interaction vectors \cite{Nagaosa2013, Fert2017, Pylypovskyi2018}. In contrast, our platform accesses this broad family of configurations through global controls and Hamiltonian drive duration. The combined tunability of $\gamma$ and $Q$ establishes the versatility of this method for preparing and controlling topological textures.

\section{Spin decoherence from magnetic field fluctuations} 
\label{sec_magnetic_field_noise}

Decoherence ultimately restricts the timescale over which simulations can be performed. To determine whether magnetic-field noise has a significant effect during state initialisation, we independently characterise the system's magnetic field. Since the splitting of the energy levels of the ground-state spin states is linearly sensitive to the axial magnetic field, fluctuations in the magnetic field cause spin dephasing. Here, we quantify the effects of magnetic field fluctuations in our system. To isolate this effect, we perform a Ramsey-type spin-echo sequence to cancel out quasi-static magnetic field offsets while remaining sensitive to higher-frequency fluctuations. 

We investigate the behaviour of the order parameter under an AC magnetic field 

\begin{equation}
\label{eqn:AC_B_field}
B(t)=B\sin{\left(2\pi f(t-t_0)\right)},
\end{equation} 

in addition to the main quasi-static axial field, $B_0$. Here, $B,f,$ and $t_0$ represent the amplitude, frequency and time offset of the AC magnetic field, respectively. Assuming linear Zeeman coupling, the qubit frequency shifts as $\omega_s(t)=\omega_s+\Gamma B(t)$, and the phase accumulated in a standard Ramsey sequence is $\phi(T) = \Gamma\int_{0}^{T}B(t)dt$. By introducing a $\pi$-pulse at $T/2$, the phase accumulated in a spin-echo experiment becomes:
\begin{equation}
\label{eqn:Ramsey_spin_echo}
\phi_\mathrm{SE}(T) = \Gamma\left[\int_{0}^{T/2}B(t)dt-\int_{T/2}^{T}B(t)dt\right].
\end{equation}
Here, the phase accumulated by the two arms has opposite signs due to the inversion of the spin state. While this results in a cancellation of DC magnetic field offsets, the spin-echo sequence retains sensitivity to AC magnetic-field components with frequencies of order $1/T$. Experimentally, we determine $\phi_\mathrm{SE}$ by measuring the ensemble projections $\langle \hat\sigma_X \rangle$ and $\langle \hat\sigma_Y \rangle$ across a range of arm times $T/2$, where

\begin{equation}
\label{eqn:spin_echo_phase}
\phi_\mathrm{SE}=\mathrm{atan2}\left(\frac{\langle \hat\sigma_Y \rangle}{\langle \hat\sigma_X \rangle}\right).
\end{equation}

\begin{figure}[!htb]
  \includegraphics{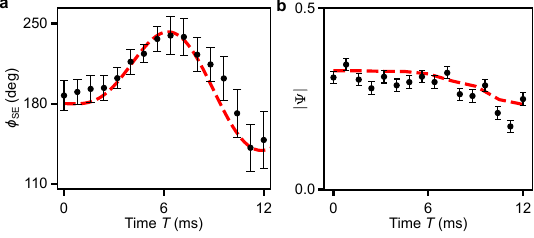}
  \centering
  \caption{\textbf{Impact of magnetic field noise on skyrmion texture.} 
  (a) Accumulated phase $\phi_\mathrm{SE}$ of the spins undergoing a Ramsey spin-echo sequence as a function of total precession time $T$. Black data points represent the mean accumulated phase reconstructed from experimental ensemble projections $\langle\hat\sigma_X\rangle$ and $\langle\hat\sigma_Y\rangle$. The red dashed line shows a fit to the spin-echo phase evolution model from Equation \eqref{eqn:Ramsey_spin_echo}, identifying a dominant \qty{100}{\hertz} magnetic field fluctuation.
  (b) Order parameter decay of the skyrmion spin texture under a Ramsey spin-echo experiment. The skyrmion state is first initialised before allowing the spins to freely precess within the trap's magnetic field for time $T/2$. Next, a global $\pi|_Y$ pulse is applied and spins precess further for time $T/2$ before being projected along $\mathcal{X},\mathcal{Y},$ and $\mathcal{Z}$. The experimental results (black points) show the order parameter retaining $\approx73\%$ of its initial value after \qty{12}{\milli\second}. This is in close agreement with the simulated time evolution (red dashed line), suggesting that the identified \qty{100}{\hertz} component is a primary source of decoherence on this timescale.}
  \label{supp_fig2:order_parameter_dephasing}
\end{figure}

Here, $\mathrm{atan2}$ is the four-quadrant arctangent, which preserves the sign of both $\langle \hat\sigma_X \rangle$ and $\langle \hat\sigma_Y \rangle$ for an unambiguous determination of the accumulated phase $\phi_\mathrm{SE}$. To characterise the noise environment, we model the phase accumulation by substituting Equation \eqref{eqn:AC_B_field} into Equation \eqref{eqn:Ramsey_spin_echo}. We fit $B$, $f$, and $t_0$ to the measured phase accumulation. Restricting the fit to a single frequency, we identify a dominant \qty{100}{\hertz} magnetic-field modulation of a few nanotesla (see Figure \ref{supp_fig2:order_parameter_dephasing}a). 

We evaluate the impact of this noise on our system by comparing the experimentally measured decay of the order parameter magnitude with a simulation based on the fitted noise model. Experimentally, we prepare the skyrmion spin texture and perform a Ramsey spin-echo experiment with a global $\pi|_Y$ pulse applied halfway through the evolution. We vary the total spin-echo evolution time, and, as in the main text, measure the $\hat\sigma_j^Y$ and $\hat\sigma_j^Z$ projections of each ion to calculate the order parameter magnitude using Equation 6 of the main text. 

For the theoretical evolution, we initialise a simulated crystal with ion positions obtained from experimentally measured coordinates. The spin configuration of each ion is initialised according to Equation 4 of the main text with $\Omega_Rt=\pi$, resulting in the formation of an ideal skyrmion texture. We then evolve this initial state by applying the experimentally measured spin-echo phase accumulation, $\phi_\mathrm{SE}$, to the spin texture and calculate the corresponding order parameter. In this way, the measured magnetic-field fluctuations are mapped onto the evolution of the initially prepared skyrmion texture without introducing any additional dephasing. As shown in Figure \ref{supp_fig2:order_parameter_dephasing}b, the experimental order parameter retains about 73\% of its initial value after \qty{12}{\milli\second}, a result that is well captured by the simulated noise evolution. While the magnetic-field fluctuations produce noticeable dephasing over millisecond timescales, no appreciable decay of the order parameter is observed over the sub-millisecond timescale relevant for state initialisation. We therefore do not include this noise source in the theoretical modelling of the order parameter during state initialisation. The observed millisecond-scale decay is consistent with a contribution from magnetic-field fluctuations associated with the identified $\qty{100}{\hertz}$ component. Potential mitigation strategies include active cancellation via out-of-phase sinusoidal currents applied to the trap's room-temperature shim coils, or the implementation of robust dynamical decoupling sequences.

\bibliography{sn-bibliography}

\end{document}